\documentclass[12pt]{article}
\usepackage{epsfig}
\usepackage{rotating}
\def\br(#1,#2){\left\langle#1#2\right\rangle}
\def\sq(#1,#2){\left[#1#2\right]}
\def\s(#1,#2){s_{#1 #2}}
\def\t(#1,#2,#3){s_{#1 #2 #3}}

\begin{document}
\begin{titlepage}

\hspace*{\fill}\parbox[t]{5cm} {\today} \vskip2cm

\begin{center}
{\Large \bf top2008 Conference Summary} \\
\medskip
\bigskip\bigskip\bigskip\bigskip
{\large {\bf S.~Willenbrock}} \\
\bigskip\bigskip\medskip

Department of Physics, University of Illinois at Urbana-Champaign \\
1110 West Green Street, Urbana, IL\ \ 61801 \\ \bigskip
\end{center}

\bigskip\bigskip\bigskip

\begin{abstract}
This is a summary of the talks presented at the International
Workshop on Top Quark Physics (top2008) held in Elba, Italy, May
18-24, 2008.
\end{abstract}

\end{titlepage}

\section{Introduction}\label{sec:intro}

We heard 48 talks at this workshop, consisting of 1571 slides, and I
am supposed to summarize it all in 60 minutes.  I will emphasize the
Tevatron, since it is producing all the top-quark physics results
right now. The LHC will soon have its day, and will eventually
dominate this conference series.  I will mostly refer to the talks
at the workshop (in parentheses), where a more complete set of
references can be found.

On April 1, 2008, Fermilab announced that it has revised the
Tevatron schedule so that the next shutdown will occur in Spring
2009.  This would allow the machine to run until the end of FY2010
with just a single shutdown.  Although the Tevatron is currently
approved to run only until the end of FY2009, the laboratory has
requested a one-year extension.  A decision on this request has not
yet been made.  Including the extension, an integrated luminosity of
over 8 fb$^{-1}$ could potentially be delivered to the experiments
(Glenzinski).

The LHC will begin operations later this year, and could deliver on
the order of 10 pb$^{-1}$ to the experiments at a collider energy of
10 TeV.  The expectation is that the collider energy will be
increased to the full 14 TeV in 2009, and will deliver an integrated
luminosity on the order of 1 fb$^{-1}$.  There is still a lot of
uncertainty in the schedule, as we are reminded by the recent
problem with the plug-in modules that connect the dipole magnets
together (Bailey).  That problem has been solved, but there may be
more bumps in the road.  The important thing is that the LHC
eventually works, which will be an amazing feat.

ATLAS and CMS have made tremendous progress, and are getting ready
to take their first data (Acosta, Schiavi).  It is amazing to see
images of the CMS cavern, empty just two years ago, now filled with
the enormous detector.  Cosmic ray data whets our appetite for the
first collider data.

\section{$t\bar t$ cross section}\label{sec:tt}

The best single measurement of the $t\bar t$ cross section comes
from the lepton plus jets channel with one or more $b$ tags (Castro,
Sharyy),
\begin{eqnarray}
\sigma(t\bar t) &=& 8.2 \pm 0.5 \,({\rm stat}) \pm 0.8 \,({\rm sys})
\pm 0.5 \,({\rm lum})
\,{\rm pb} \;({\rm CDF}) \nonumber \\
\sigma(t\bar t) &=& 8.1 \pm 0.5 \,({\rm stat}) \pm 0.7 \,({\rm sys})
\pm 0.5 \,({\rm lum}) \,{\rm pb} \;({\rm D0}) \nonumber.
\end{eqnarray}
The D0 experiment has also made a measurement of the cross section
in the $\tau +$lepton and $\tau +$jets channels (Sharyy). Although
it is not as accurate as the best measurements, it is impressive
that such a measurement can be made.  Keep in mind that a little
over 20 years ago, the tau lepton was one of the sources of the
``monojets'' (not to be confused with the top-quark monojets
referred to by Chevallier) seen by the UA1 experiment
\cite{Arnison:1984qu}.  To quote the abstract of that paper, ``We
report the observation of five events in which a missing transverse
energy larger than 40 GeV is associated with a narrow hadronic jet
...'' We now recognize this as $W\to \tau\nu$, but at the time it
was thought that it might be new physics.  It is difficult to
predict what things will puzzle us when we begin the operation of
the LHC, but it is hard to imagine that there won't be some
confusion.

The measurements of the $t\bar t$ cross section are in good
agreement with the predictions of next-to-leading-order (NLO) QCD.
Theorists have been working hard to go beyond NLO, and to estimate
the uncertainties in the prediction (Mangano).  The uncertainty from
the parton distribution functions (PDF's) is quite small, around 6\%
for the Tevatron and 3\% for the LHC according to CTEQ6.5, but it is
perplexing that MRSTW-06 gives an uncertainty about half that
amount.  Even worse, the central values of the cross section
calculated from the two PDF's differ by about 6\% at the LHC, which
is nearly twice the uncertainty from the PDF's.  This needs to be
sorted out.

Theorists are making progress towards a full NNLO calculation of the
$t\bar t$ cross section, but there is still a ways to go (Mangano).
In the meanwhile, there exist partial NNLO calculations.  One recent
calculation shows greatly reduced scale dependence, but this might
be an artifact of equating the renormalization and factorization
scales \cite{Moch:2008qy}.  Recall that the former has to do with
ultraviolet physics while the latter is related to collinear
physics, so they are logically independent.

I would like to raise a separate issue, which is whether we should
attach any significance to scale dependence at all, in particular
whether we should use it to estimate theoretical uncertainties
\cite{Maltoni:2007tc}. Consider a very basic process, $Z$ production
at the Tevatron.  At leading order (LO), the cross section is
predicted to be around 5.8 nb, with almost no scale dependence. The
NLO cross section is about 7 nb, well outside of the range of the LO
scale dependence; it also has almost no scale dependence.  It agrees
well with the NNLO cross section, as well as with the measured cross
section.  This suggests that one should take the uncertainty
associated with scale dependence with a grain of salt, since it is
misleading even in this very basic case.

Theorists have also calculated the NLO electroweak corrections to
the $t\bar t$ cross section, and they are generally very small (Si).
However, the corrections becomes significant, of order 10\%, when
considering top quarks at transverse momenta greater than 1 TeV at
the LHC.

The cross section for $t\bar t+1$ jet has also been calculated at
NLO, and shows the usual reduction in scale dependence compared with
LO (Uwer).  The forward-backward asymmetry (also present in
inclusive $t\bar t$ production \cite{Kuhn:1998jr}) at the Tevatron
shows a large NLO correction, well outside the scale uncertainty of
the LO calculation.  Given my remarks above, I am not concerned
about this, but the authors are currently considering whether there
is an observable that is more stable under radiative corrections.

The LHC will be a top factory, and a top-quark signal is expected
already in the first 10 pb$^{-1}$ (Cobal and Tsirigkas).  It is
anticipated that the $t\bar t$ cross section will be measured with
an uncertainty of 5-10\% (systematics dominated) in the first 100
pb$^{-1}$.

\section{Top-quark mass}\label{sec:mt}

The top-quark mass is now known with remarkable precision, $m_t =
172.6 \pm 1.4$ GeV.  I consider it a great success that the central
value of the measured mass has not moved very much from its present
value, going all the way back to the first measurement
\cite{Abe:1994st}.  To convince you that this is nontrivial,
consider the case of the third-generation charged lepton, the tau,
whose mass has moved significantly over the years \cite{Yao:2006px}.
The consistency of the measured mass with the indirect mass
extracted from precision electroweak data is another remarkable
achievement, and gives us great confidence in the standard
electroweak theory.

\begin{figure}[ht]
\begin{center}
\vspace*{.2cm} \hspace*{0cm} \epsfxsize=14cm \epsfbox{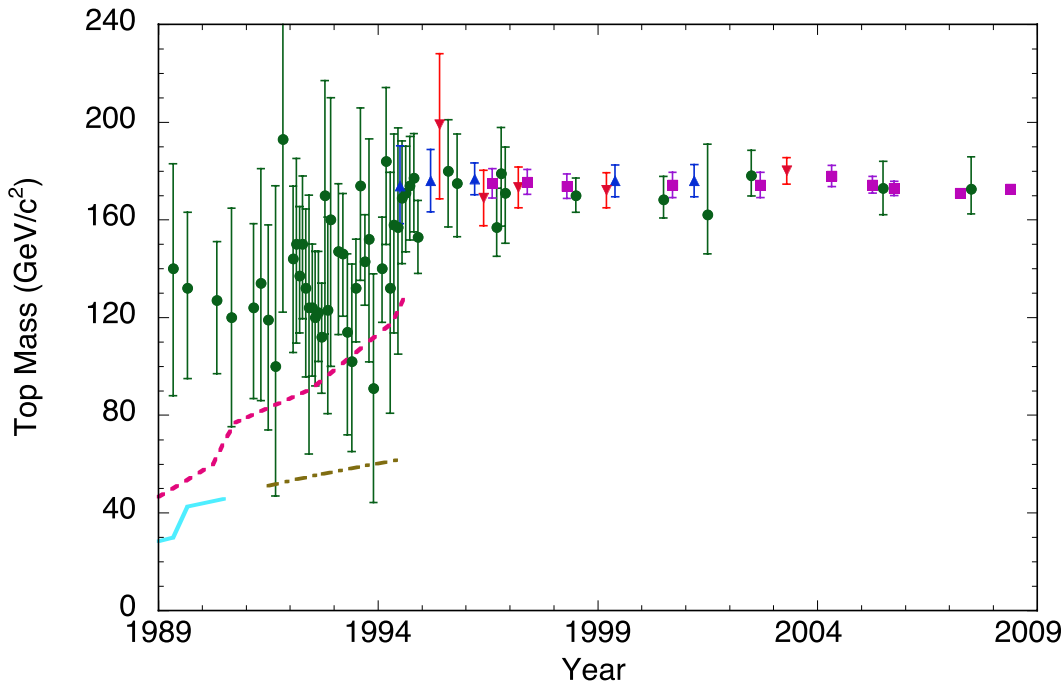}
\vspace*{-.8cm}
\end{center}
\caption{The central value of the top-quark mass has remained
remarkably stable over the years, while the uncertainty has
decreased dramatically \cite{Quigg:1997uh}.} \label{fig:TopMass2008}
\end{figure}

There was quite a bit of discussion of the top-quark mass definition
at the workshop, so I'd like to pause here to review what we mean by
a quark pole mass (Glenzinski).  The propagation of a physical
particle, such as an electron, is described in quantum field theory
by a propagator, which is proportional to $1/(p^2-m^2)$, where $p$
is the particle's four-momentum.  This propagator has a simple pole,
in the language of complex analysis, at $p^2 = m^2$, and hence this
mass is called the pole mass; for the electron, its value is
$m=0.511$ MeV. A quark propagator is similarly described.  However,
quarks are always confined, so there is really no such thing as a
freely propagating quark, and hence the pole mass is not really
physical. Since the scale of confinement is $\Lambda_{\rm
QCD}\approx 200$ MeV, the pole mass is only defined up to an
ambiguity of order $\Lambda_{\rm QCD}$.  Thus we arrive at a
definition of a quark pole mass: it's the mass the quark would have
in the absence of confinement. Unfortunately, we cannot simply turn
off confinement, so a quark pole mass is inherently ambiguous (of
order $\Lambda_{\rm QCD}$).

In all the calculations that are used to extract the top-quark mass
from the experimental data, the top-quark propagator is described in
the standard way (also including the width of the top quark).  Hence
the top-quark mass that we extract from experiment is the pole mass.
Since the present uncertainty is large compared with $\Lambda_{\rm
QCD}$, we ignore the inherent ambiguity in the pole mass.

An idealized study suggests that the dependence of the top-quark
mass extracted at the Tevatron on the hadronization (confinement)
model has a non-perturbative component that is about 500 MeV,
consistent with an ambiguity of order $\Lambda_{\rm QCD}$ in the
pole mass (Wicke).

As with the $t\bar t$ cross section, the best measurements of the
top-quark mass come from the lepton plus jets channel (van Remortel,
Renkel), \newpage
\begin{eqnarray}
m_t &=& 171.4 \pm 1.5 \,({\rm stat+JES}) \pm 1.0 \,({\rm sys}) \,{\rm GeV} \;({\rm CDF}) \nonumber \\
m_t &=& 172.2 \pm 1.1 \,({\rm stat}) \pm 1.6 \,({\rm sys+JES})
\,{\rm GeV} \;({\rm D0}) \nonumber
\end{eqnarray}
where JES = Jet Energy Scale.  The combined CDF and D0 mass
measurement, $m_t=172.6 \pm 1.4$ GeV, has an uncertainty that is
approaching $\Lambda_{\rm QCD}$, and could reach 1.0 GeV by the end
of Run II (Glenzinski).  At the LHC, the statistical uncertainty
will be negligible, and the signal/background ratio is significantly
greater than at the Tevatron, so one could contemplate a measurement
with an accuracy of even less than 1.0 GeV (Sjolin and Wolf). Along
with accurate knowledge of the jet energy scale, this would require
taking the effects of confinement into account quantitatively.

There is no reason in principle that this could not be done.  In
fact, it has already been achieved for the proposed ILC, where a
scan of the $t\bar t$ threshold could be used to measure the
top-quark mass to an accuracy of 75 MeV (Juste).  This is not the
pole mass, but rather a short-distance mass that is free of the
ambiguity of order $\Lambda_{\rm QCD}$ that plagues the pole mass.

Recently attention has turned to measuring the top-quark mass at the
ILC above the $t\bar t$ threshold (Hoang).  This is closer in spirit
to the situation at the Tevatron and the LHC, although it is simpler
since the initial state is colorless.  Using a series of effective
field theories, it has been shown how to isolate the nonperturbative
contribution to the top-quark mass into a universal soft function
that can be extracted from data.  Perhaps something similar can be
developed for hadron colliders.

I would like to speculate that one might be able to do even better.
The top quark is produced on a very short time scale, of order
$m_t^{-1}$, and propagates for a time scale of order $1/\Gamma_t$
before it decays to a $W$ boson and a $b$ quark. Only later, on a
time scale of order $1/\Lambda_{\rm QCD}$, does the process of
hadronization (confinement) of the $b$ quark take place. That means
that the $W$ boson is emitted well before the effects of confinement
are felt. Thus

\begin{figure}[hb!]
\begin{center}
\vspace*{.2cm} \hspace*{0cm} \epsfxsize=14cm \epsfbox{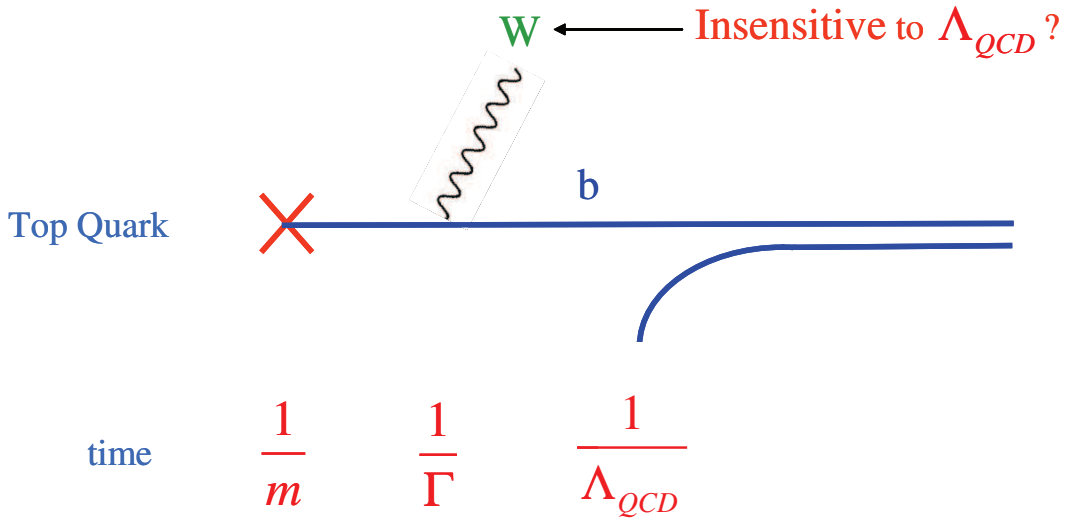}
\vspace*{-.8cm}
\end{center}
\caption{The top quark decays before it feels the effects of
nonperturbative QCD (confinement).} \label{fig:topdecay}
\end{figure}

\noindent the decay products of the $W$ boson may be insensitive to
the effects of confinement, and by studying them we might be able to
extract the top-quark mass with less dependence on the quantitative
details of confinement \cite{Bigi:1986jk}. Indeed, there are studies
of the top-quark mass from the lepton $p_T$ spectrum (Giokaris), or
from the lepton plus $J/\psi\to \ell^+\ell^-$ spectrum (Sjolin and
Wolf), at the LHC. Perhaps these approaches will prove to be
fruitful to reduce the quantitative effects of confinement.

Incidentally, this argument also demonstrates why it is not possible
to avoid the ambiguity in the top-quark pole mass of order
$\Lambda_{\rm QCD}$, even though the top quark decays before it
hadronizes.  The top-quark mass is reconstructed from the invariant
mass of the $W$ boson and the $b$ quark, but because the $b$ quark
hadronizes, this invariant mass is unavoidably ambiguous (of order
$\Lambda_{\rm QCD}$) \cite{Smith:1996xz}.

\section{Single top}\label{sec:t}

\begin{figure}[b]
\begin{center}
\vspace*{.2cm} \hspace*{0cm} \epsfxsize=14cm \epsfbox{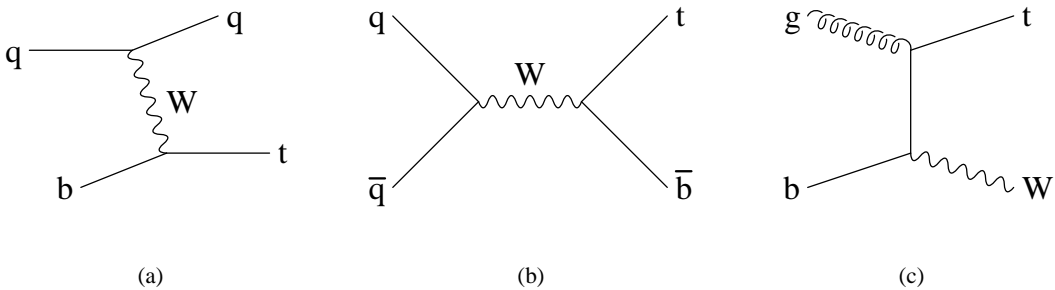}
\vspace*{-.8cm}
\end{center}
\caption{Single top quarks are produced via the $t$-channel,
$s$-channel, and $Wt$ associated production processes.}
\label{fig:singletop}
\end{figure}

Single top quarks are produced via the weak interactions via three
processes, dubbed $s$-channel, $t$-channel, and $Wt$ associated
production (Jabeen, Lueck).  While $t\bar t$ production can easily
be separated from backgrounds in the lepton plus four jets mode
(with one or more $b$ tags), distinguishing single top production in
the lepton plus two jets mode (with one or more $b$ tags) requires a
more sophisticated analysis, because the backgrounds are much more
severe. It is amazing that the separation can be done at all. Both
CDF and D0 have made measurements of the single-top cross section,
\begin{eqnarray}
\sigma(t) &=& 2.2 \pm 0.7 \,{\rm pb} \;({\rm CDF}) \nonumber \\
\sigma(t) &=& 4.7 \pm 1.3 \,{\rm pb} \;({\rm D0}) \nonumber
\end{eqnarray}
in good agreement with the standard-model prediction with
$|V_{tb}|=1$.  Both experiments have made a first attempt to
separate the $s$-channel and $t$-channel signals (Lueck), something
we will see more of as the integrated luminosity mounts.  Using
their cross-section measurements, both experiments have extracted a
value for $|V_{tb}|$,
\begin{eqnarray}
|V_{tb}| &=& 0.89 \pm 0.14 \,({\rm exp}) \pm 0.07 \,({\rm theory})\; ({\rm CDF}) \nonumber \\
{|V_{tb}|}2 &=& 1.72^{+0.64}_{-0.54}\; ({\rm D0}). \nonumber
\end{eqnarray}

Ideally this information will be incorporated into a global fit of
the CKM matrix.  In this sense, we are now members of the CKM
family.  However, we need to think about what assumptions we are
making. Since single-top production is dominated by the $t$-channel
process, we usually assume that the initial parton is a $b$ quark
that gets converted into a top quark via the weak interaction, with
a rate proportional to $|V_{tb}|^2$. However, it is also possible
that the initial quark is a $s$ or $d$ quark, in which case the rate
is proportional to $|V_{ts}|^2$ or $|V_{td}|^2$, respectively
(Frederix). Thus we can use the measured cross section to constrain
the parameter space of $|V_{tb}|, |V_{ts}|, |V_{td}|$.  This is in
the same spirit as constraining the $\bar\rho, \bar\eta$ plane
(Bucci).

Despite this, one often wants to just quote a value for $|V_{tb}|$,
and it is important to consider the minimum set of assumptions that
allows one to do this.  One could say that the assumption is that
there are just three generations. While there is nothing wrong with
that, there is a weaker assumption that one can make.  If we simply
assume that $|V_{tb}| >> |V_{ts}|,|V_{td}|$ (with no assumptions
about the number of generations), then single-top production is
dominated by the $t$-channel process with an initial $b$ quark,
which is proportional to $|V_{tb}|^2$.

Theorists have been working hard to make accurate predictions for
single-top production.  All three processes have been calculated at
NLO, including decays of the top quark and spin correlations
(Tramontano).  Electroweak corrections have also been calculated
and, as in the case of $t\bar t$ production, the corrections are
very small except for high-$p_T$ top quarks, in which case the
corrections are of order 10\% at the LHC (Mirabella). The idea that
one could observe single-top photoproduction at the LHC was
discussed (Ovyn).

Single-top production will be observable at the LHC with around 10
fb$^{-1}$, including the first observation of $Wt$ associated
production.  The $s$-channel process will be the most challenging to
observe at the LHC (Cristinziani and Petrucciani).

\section{Anomalous couplings}\label{sec:anomalous}

We heard a great deal about anomalous top-quark weak interactions at
this workshop (Aguilar-Saavedra, Brandt, Jabeen, Juste, Shabalina,
Spiga, Veloso).  These are usually parameterized in terms of a
vertex function that describes the $Wtb$ vertex, assuming the top
and bottom quarks are on shell;
\begin{eqnarray}
\Gamma^\mu & = & -\frac{g}{\sqrt
2}V_{tb}\left({\gamma^\mu[f_1^LP_L+f_1^RP_R]-\frac{i\sigma^{\mu\nu}}{M_W}(p_t-p_b)_\nu
[f_2^LP_L+f_2^RP_R]}\right) \nonumber
\end{eqnarray}
where $P_{L,R} = (1\mp \gamma_5)/2$.  This is the most general
vertex function consistent with Lorentz invariance for on-shell top
and bottom quarks.  The $W$ boson may be off-shell with virtuality
$p_W^2$, and the form factors $f_{1,2}^{L,R}$ are unknown functions
of $p_W^2$.  If we allow the top and bottom quarks to be off shell,
the vertex function becomes much more complicated
(Aguilar-Saavedra). There are sixteen additional form factors, and
all form factors are unknown functions of $p_W^2,p_t^2,p_b^2$.

There is an alternative approach based on the concept of an
effective field theory \cite{Cao:2007ea}.  In this approach, one
accepts the standard electroweak theory as the correct zeroth-order
approximation, and parameterizes physics beyond the standard model
in terms of higher-dimension operators, suppressed by inverse powers
of a large mass scale $\Lambda >> M_W$.  The leading
higher-dimension operators are those of the lowest dimensionality,
since they are suppressed by the least inverse powers of $\Lambda$.
These leading operators are of dimension six, so the effective
Lagrangian is
\begin{eqnarray}
{\cal L} & = & {\cal L}_{SM} + \frac{1}{\Lambda^2}\sum_i(c_i{\cal
O}_i+h.c.) \nonumber
\end{eqnarray}
where ${\cal O}_i$ are the dimension-six operators.  There are many
such operators, but only three of them are expected to be
significant if the physics beyond the standard model is a
weakly-coupled gauge theory, and one of those three is already very
constrained by $b\to s\gamma$.  By comparing data with theory one
can put upper bounds on the coefficients $c_i/\Lambda^2$. Although
this has not been the approach that has usually been used for
anomalous couplings, I think it is a lot cleaner, and I believe we
will be seeing more of it in the future.

\section{Tools for top}\label{sec:tools}

The number of tools available to us for top-quark physics (and
collider physics in general) has grown enormously over the years,
and they have become ever more sophisticated (Maltoni).  There are
three broad classifications of codes that use parton showers (PS):
Matrix Element plus PS (Herwig, Pythia, etc.); Matrix Element plus
PS plus Merging (Alpgen, MadGraph, Sherpa); and NLO plus PS (MC@NLO,
POWHEG).  There is no code that includes NLO plus PS plus Merging,
although that is probably within reach.  These codes are incredibly
useful for comparing theory with experiment.

Although these codes are generally accurate, they are not perfect,
and by comparing different codes we can appreciate their strengths
and weaknesses, and hopefully improve them.  One example relevant to
top-quark physics is the rapidity distribution of the hardest jet
(not from top decay) in $t\bar t+$jets events.  There is a
significant discrepancy between Herwig and Alpgen, and also between
Pythia and MadGraph, for jets at central rapidity, and the
discrepancy becomes greater as the jet $p_T$ increases.  Another
example is the $p_T$ of the second $b$ quark (not from top decay) in
$t$-channel single-top production.  This second $b$ arises from
initial-state gluon splitting, $g\to b\bar b$; one $b$ is
transformed into a top quark by a $t$-channel $W$, and the other
tends to reside at low $p_T$.  The $p_T$ and rapidity distributions
of this second $b$ differ significantly between a pure
Matrix-Element calculation and a Matrix-Element plus PS calculation
(Maltoni). This is dealt with in the best way we know how for
single-top at the Tevatron, but we should be able to do better
(Huseman).  The same issue is present at the LHC (Chierici).

An area with a lot of current activity, that will also be important
for the LHC, is $W,Z$ production in association with heavy quarks
($c,b$).  The first measurements from CDF and D0 of $W+c$-jet are in
rough agreement with NLO QCD, and it will be interesting to see how
this evolves with more data (Harel).  The CDF measurement of
$Z+b$-jet agrees well with NLO QCD at high $p_T$, but exceeds theory
at low $p_T$ (Glenzinski).

Given the large statistics for top-quark physics at the LHC, most
measurements will be systematics limited.  One idea to improve the
jet energy scale at the LHC is to use the top-quark mass measured at
the Tevatron as a constraint (Van Mulders and Bachacou).  This is
truly ``TeV4LHC,'' in the spirit of the workshop held in 2004-2005
\cite{Gerber:2007xk}.

\section{New ideas for top}\label{sec:new}

The previous sections already contain some new ideas for top, so
this section is meant to capture ideas that didn't fit neatly
elsewhere.

There are several theories beyond the standard model in which
top-quark physics plays a special role.  Those who believe in the
standard Higgs model of electroweak symmetry breaking were accused
of being ``conventional,'' a label that no particle physicist likes
(Holdom).  An alternative is to have the electroweak symmetry broken
by a fourth generation of fermions and a new, strongly-interacting
gauge interaction that acts on the third and fourth generations.
This gives rise to both a heavy $t^\prime$ quark and a heavy $X$
gauge boson that decays to $t\bar t$.  A heavy object decaying to
$t\bar t$ appears in many extensions of the standard model, and
would manifest itself as a resonance in the $t\bar t$ invariant mass
spectrum (Tait).  The decay $t^\prime \to Wq$ has been sought at the
Tevatron, and a lower bound of 284 GeV (95\% CL) has been placed on
the $t^\prime$ mass by CDF (Sorin).

A classic standard-model effect in $t\bar t$ production that has yet
to be confirmed is the correlation between the spins of the $t$ and
$\bar t$ (Stal, Spiga, Veloso).  The correlation differs between
$q\bar q\to t\bar t$ and $gg\to t\bar t$, so a measurement of the
spin correlation would allow one to extract the fraction of events
from each production mechanism.  Another method to extract this
fraction is to study the underlying activity, which is expected to
be greater for processes initiated by gluons.  Using this technique,
CDF has measured the fraction of events from $gg\to t\bar t$ to be
consistent with zero (Pashapour).

Another standard-model effect that is being sought is the $t\bar t$
charge asymmetry (Pasha\-pour, Shabalina).  This asymmetry, present
only for $q\bar q\to t\bar t$, is a difference between the rate at
which top quarks and top antiquarks are produced at a given angle
with respect to the incoming quark parton.  It is a small effect
because it arises first at NLO in QCD.  At the Tevatron, it
manifests itself as a forward-backward asymmetry of top quarks
produced with respect to the proton beam.  A recent measurement from
D0,
\begin{eqnarray}
A & = & (12\pm 8\,({\rm stat})\pm 1\,({\rm syst}))\%\; ({\rm D0})
\nonumber
\end{eqnarray}
made in the $W+\ge 4$ jets mode, is consistent with the theoretical
prediction of $4-5\%$ \cite{Kuhn:1998jr}.

Another asymmetry of interest is associated with the rapidity
distribution of $W^+$ bosons, which are produced more in the proton
direction than in the antiproton direction (Glenzinski).  The
agreement between theory and experiment is impressive, all the more
so because this is one of the few places where we have a NNLO
prediction as well as data of similar accuracy.  Although this is
often referred to as a charge asymmetry, it is of a completely
different origin than the $t\bar t$ charge asymmetry discussed
above, and is merely a consequence of the parton distribution
functions. I find it preferable to call it a $W$ rapidity asymmetry
\cite{Berger:1988tu}.

The production of a Higgs particle in association with $t\bar t$ at
the LHC would give us a direct measurement of the top-quark Yukawa
coupling to the Higgs (Aad and Steggemann).  Followed by the decay
$h\to b\bar b$, this signal is very challenging to extract from the
backgrounds, but its importance makes it worth the effort.  Even
harder is to extract a signal for Higgs production in association
with single top \cite{Maltoni:2001hu}.  Perhaps one should attempt
that first; then $t\bar th$ will seem relatively easy!

We heard a variety of talks on searching for exotic top-quark
physics at the Tevatron and the LHC (Sorin, Wicke, Yumiceva,
Chevallier, Blekman and Milosavijevic).  Along with $X\to t\bar t$
and $t^\prime$ mentioned above, there was discussion of top decay
via a charged Higgs boson ($t\to h^+b$), flavor-changing
neutral-current top decay ($t\to Zq,\gamma q$), and top squarks. We
don't know which, if any, of these are realized in nature, but it
does point out that we may very well find physics beyond the
standard model by studying the top quark.  We shall see.

\section{Conclusion}\label{sec:conclusion}

This is an exciting time for top-quark physics, with data pouring
out of the Tevatron experiments.  We have seen a lot of new
top-quark physics results over the past few years, most of which
were presented at this workshop.  The next few years will witness a
continuation of the Tevatron program and the beginning of the LHC
era.  As exciting a time as it is, the best is yet to come!

\section*{Acknowledgments}

\indent\indent I am grateful for conversations with many of the
workshop participants, and especially to Fabio Maltoni for his
advice and assistance.  This work was supported in part by the
U.~S.~Department of Energy under contract No.~DE-FG02-91ER40677.


\end{document}